\documentclass{ws-rv9x6}
\usepackage{subfigure}   
\usepackage{ws-rv-thm}   
\usepackage{ws-rv-van}   
\makeindex
\usepackage{bbold}

\usepackage{ulem}
\usepackage[usenames]{xcolor}

\begin{document}

\chapter[Foundations of strangeness nuclear physics]{Foundations of Strangeness Nuclear Physics\\
                derived from chiral Effective Field Theory}\label{ra_ch1}

\author[]{Ulf-G. Mei{\ss}ner}

\address{HISKP and BCTP, Universit\"at Bonn, D-53115 Bonn, Germany \\
IKP-3, IAS-4, JARA-HPC and JARA-FAME,\\ Forschungszentrum J\"ulich,
D-52425 J\"ulich, Germany\\
meissner@hiskp.uni-bonn.de}

\author[]{Johann Haidenbauer}

\address{IKP-3 and IAS-4, Forschungszentrum J\"ulich, 
D-52425 J\"ulich, Germany\\
j.haidenbauer@fz-juelich.de}

\begin{abstract}
Dense compact objects like neutron stars or black holes have always
been one of Gerry Brown's favorite research topics. This is closely
related to the effects of strangeness in nuclear physics. Here,  
we review the chiral Effective Field Theory approach to interactions involving 
nucleons and hyperons, the possible existence of strange dibaryons,
the fate of hyperons in nuclear matter and the
present status of three-body forces involving hyperons and nucleons.
\end{abstract}
\body


\section{Introduction}\label{sec1}

Gerry Brown was a fascinating scholar and human being with a broad range
of physics interests. One of his prime foci was the physics of dense and
compact objects like neutron stars and black holes, 
see e.g. Refs.~\cite{Maxwell:1977zz,Brown:1988ik,Brown:1993jz,Brown:1994vy,Bethe:1995hv,Lee:2001xw,Schwenk:2002fq,Lee:2005jv}. 
With the observation of two solar mass neutron stars,
this physics has {taken} a new twist, commonly referred to as the ``hyperon puzzle''.
This puzzle relates to the fact that the equation of state in the presence of hyperons
is generally too soft to allow for such heavy neutron stars. Some repulsive mechanism,
may it be related to the hyperon-nucleon or hyperon-hyperon interactions, or even
to more exotic three-body forces involving hyperons, is thus required to reconcile the
presence of hyperons within such dense and compact objects with their global properties,
see e.g. Refs.~\cite{Chatterjee:2015pua,Bombaci:2016xzl} (and references therein). 
This is were some of our recent  research comes in, namely the chiral Effective Field Theory 
(EFT) based description of the  hyperon-nucleon ($YN$)  in Sec.~\ref{sec:YN}
and the hyperon-hyperon ($YY$) interactions  in Sec.~\ref{sec:YY}.
 Another related fascinating topic is the possible existence of 
exotic states with baryon number two, formed from two hyperons, which we will also discuss 
in  Sec.~\ref{sec:eco}. Next, we discuss  the behavior of hyperons in nuclear matter in Sec.~\ref{sec:med}.
Last but not least, we consider the final frontier, namely the three-body forces
involving hyperons in Sec.~\ref{sec:3bf}.
 We are quite certain that  Gerry would have loved these developments 
and we therefore dedicate this paper to his memory.



\section{Hyperon-nucleon interactions}\label{sec:YN}

The basic ingredient in hyper-nuclear physics is the hyperon-nucleon interaction.
Conventionally, it has been studied based on meson-exchange models, however, with the
advent of successfull EFT methods for the nucleon-nucleon ($NN$) interaction 
(see Refs.~\cite{Epelbaum:2008ga,Machleidt:2011zz} for reviews), one now has 
a better and more systematic handle on these fundamental interactions, firmly rooted
in the symmetries of QCD. In contrast to the $NN$ interaction, the amount of data
on $YN$ ($Y = \Lambda, \Sigma$) scattering is scarce, about 35 data points 
supplemented by the binding energies of
a few light hyper-nuclei. Therefore, any EFT description of these data has a more
exploratory character than it is the case for the $NN$ problem. In fact, the first 
work on the $YN$ interaction in EFT was
due to Korpa et al.~\cite{Korpa:2001au}, who made use of the so-called 
KSW power counting~\cite{Kaplan:1998tg}.
In the Bonn-J\"ulich group, we already had made good experiences with the so-called 
Weinberg power counting for the $NN$ interactions. In that scheme, chiral perturbation 
theory is utilized  for calculating the
potential between two or more baryons in a systematic manner. This potential is then
used within a regularized Lippmann-Schwinger (LS) equation to generate the bound and the
scattering states, $T = V + V G_0 T$, with $T$ the T-matrix and $G_0$ the two-baryon 
propagator.  The leading order (LO) $YN\to YN$ calculations were performed by Polinder and 
the authors~\cite{Polinder:2006zh} and later, the next-to-leading order (NLO) contributions 
were analyzed in collaboration with the Munich group of 
Norbert Kaiser, Stephan Petschauer and Wolfram Weise~\cite{Haidenbauer:2013oca}. For a 
comparison of the EFT and more conventional approaches, see Ref.~\cite{Haidenbauer:2007ra}.
In what follows, we eschew models.
\begin{figure}[h]
 \centering
 \includegraphics[width=0.11002\textwidth]{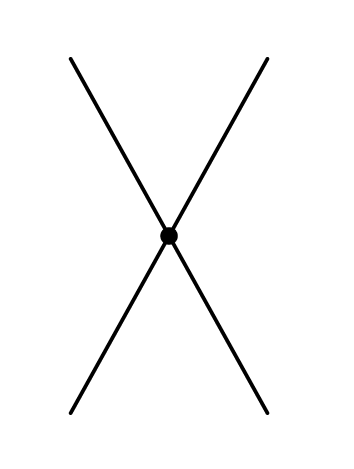}
 \includegraphics[width=0.11002\textwidth]{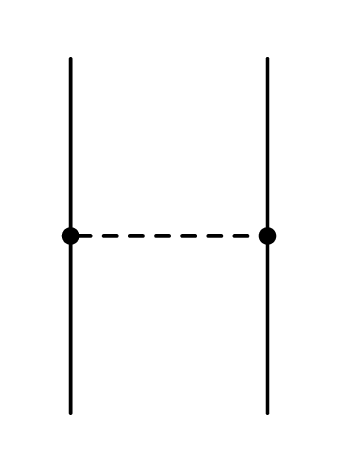}
 \includegraphics[width=0.11002\textwidth]{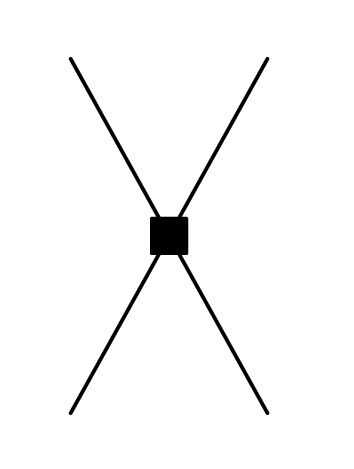}
 \includegraphics[width=0.11002\textwidth]{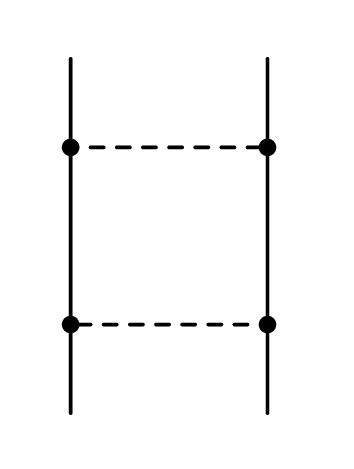}
 \includegraphics[width=0.11002\textwidth]{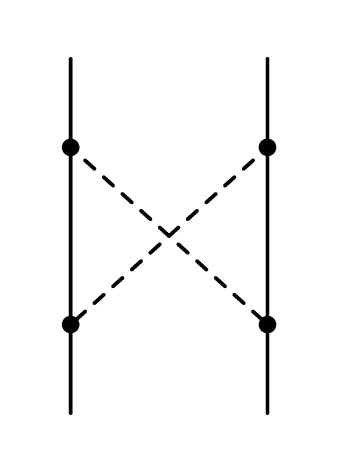}
 \includegraphics[width=0.11002\textwidth]{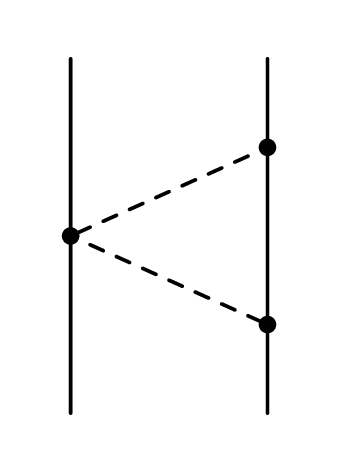}
 \includegraphics[width=0.11002\textwidth]{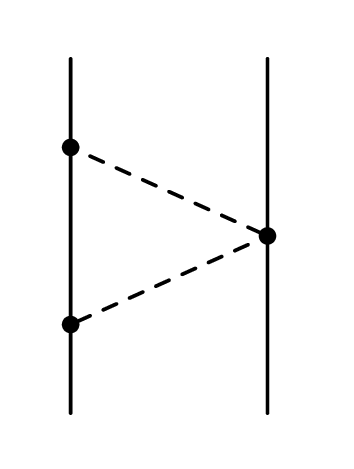}
 \includegraphics[width=0.11002\textwidth]{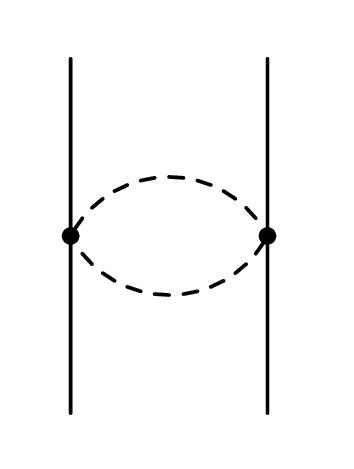}
 \caption{Relevant diagrams for the effective potential 
up-to-and-including NLO. Solid and  dashed lines denote baryons 
($N,\Lambda,\Sigma,\Xi$) and  mesons ($\pi,K,\eta$), respectively. 
The square  symbolizes a contact vertex with two derivatives.
From left to right: LO contact term, one-meson exchange, NLO contact term, planar box, 
crossed box, left triangle, right triangle, football diagram.
From the planar box graph, only the irreducible part contributes to the potential.
} 
\label{fig:dia}
\vspace{-3mm}

\end{figure}

The EFT is constructed from the asymptotically observed particles, the hadrons, here
from the baryon octet (for short, baryons) and the Goldstone-boson octet (for short,
mesons).
The basic ingredient in the EFT approach to the $YN$ interaction is the
effective potential, with its various terms ordered according to the power counting. 
At leading order, ${\mathcal O}(Q^0)$, where $Q$ denotes an external momentum or a 
Goldstone boson mass,
there are two types of contributions. First, one has one-meson-exchange graphs, with all
couplings constants expressed in terms of the pion-nucleon coupling $g_{\pi NN}$ and the
ratio of the SU(3) axial-couplings $F/(F+D)$. Second, based on group theory arguments,
there are six four-baryon contact terms without derivatives. 
From those, only five 
contribute to the scattering process $YN \to YN$ while all six occur in the $YY$ sector. 
Extending to NLO, one has now also 
contributions from two-meson-exchange diagrams and further contact interactions with 
two derivatives. The pertinent Feynamn diagrams are depicted in Fig.~\ref{fig:dia}.
All these have been calculated 
and included in the potential presented
in Ref.~\cite{Haidenbauer:2013oca}.
However, it should be stressed that some two-meson-exchanges like $K\bar K$ or $\eta K$
are so short-ranged that they 
could be also effectively absorbed in the contact terms.
The number of additional contact interactions is large, 
so in general we resort to account
only for SU(3) symmetric terms and fit only the low-energy constants (LECs) related to the
$S$-wave terms and use $P$-wave $NN$ phase shifts as constraints. 
However, physical masses of the mesons and the baryons are used throughout. 
This is needed to account properly for the fact that the pion mass is much smaller
than that of the other Goldstone bosons and also to have the correct threshold energies for 
the various baryon-baryon channels, and it introduces a certain amount of SU(3) breaking.

\begin{figure}[t]
\begin{center}
\includegraphics[width=0.32\textwidth]{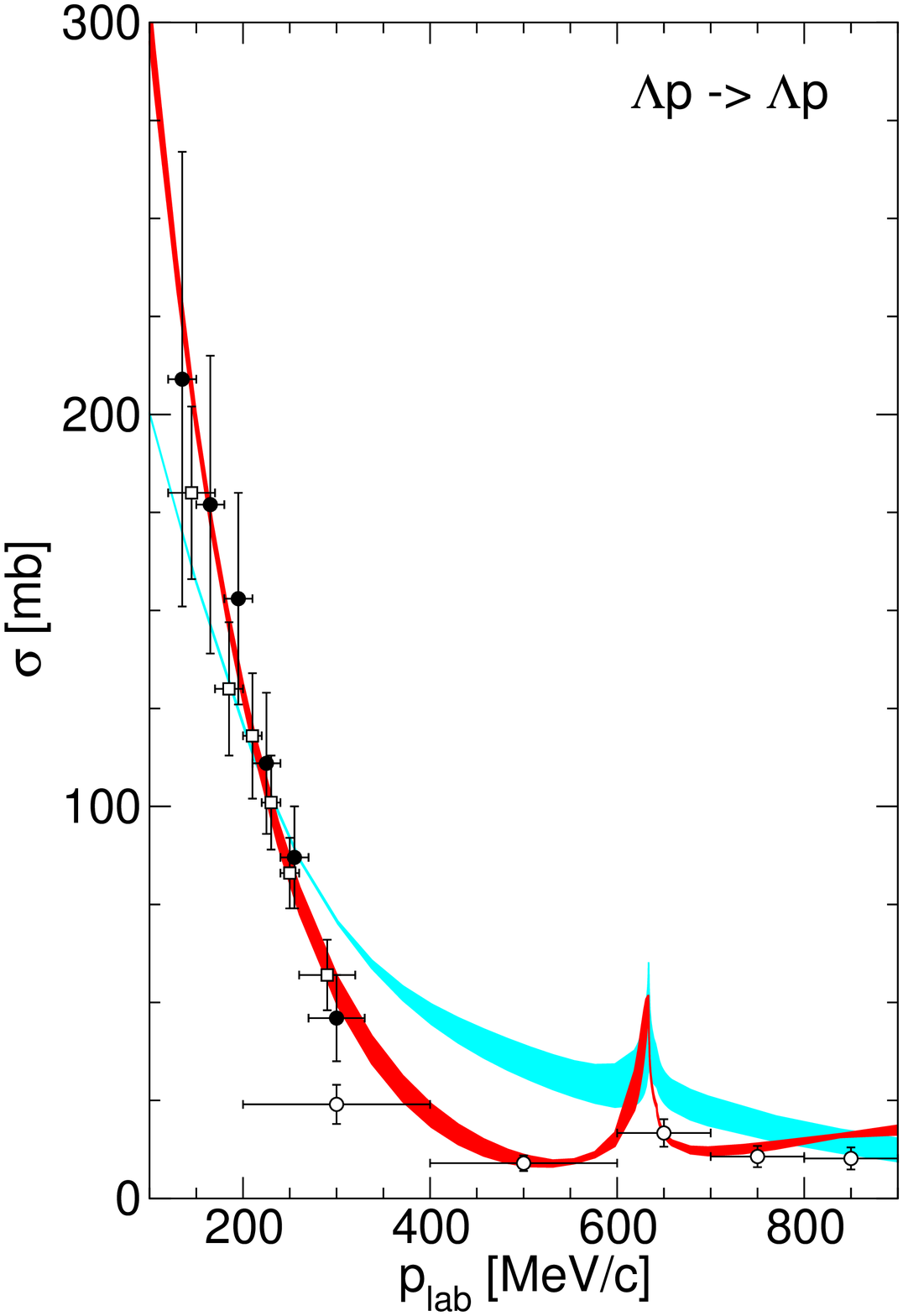}
\includegraphics[width=0.32\textwidth]{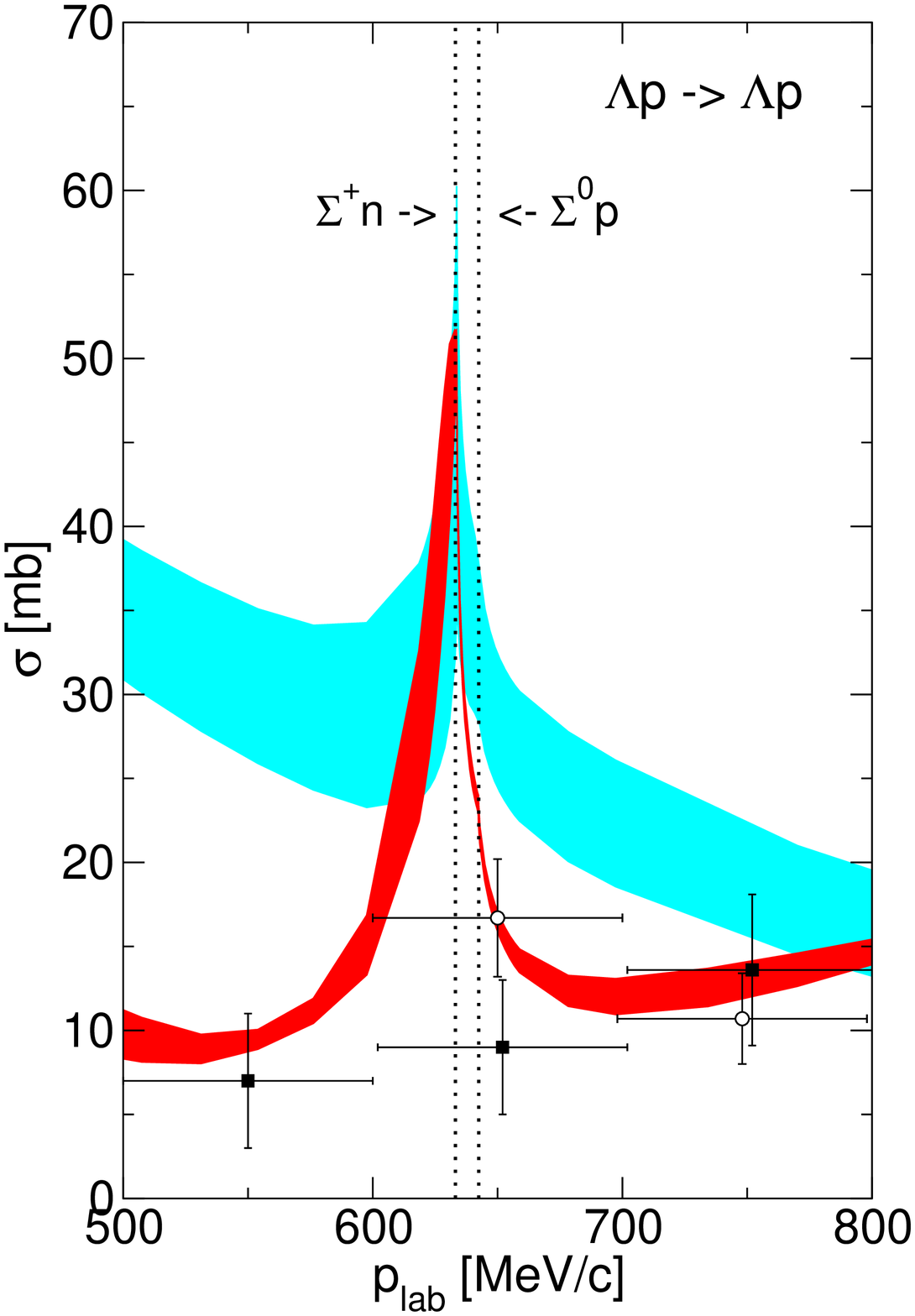}
\includegraphics[width=0.32\textwidth]{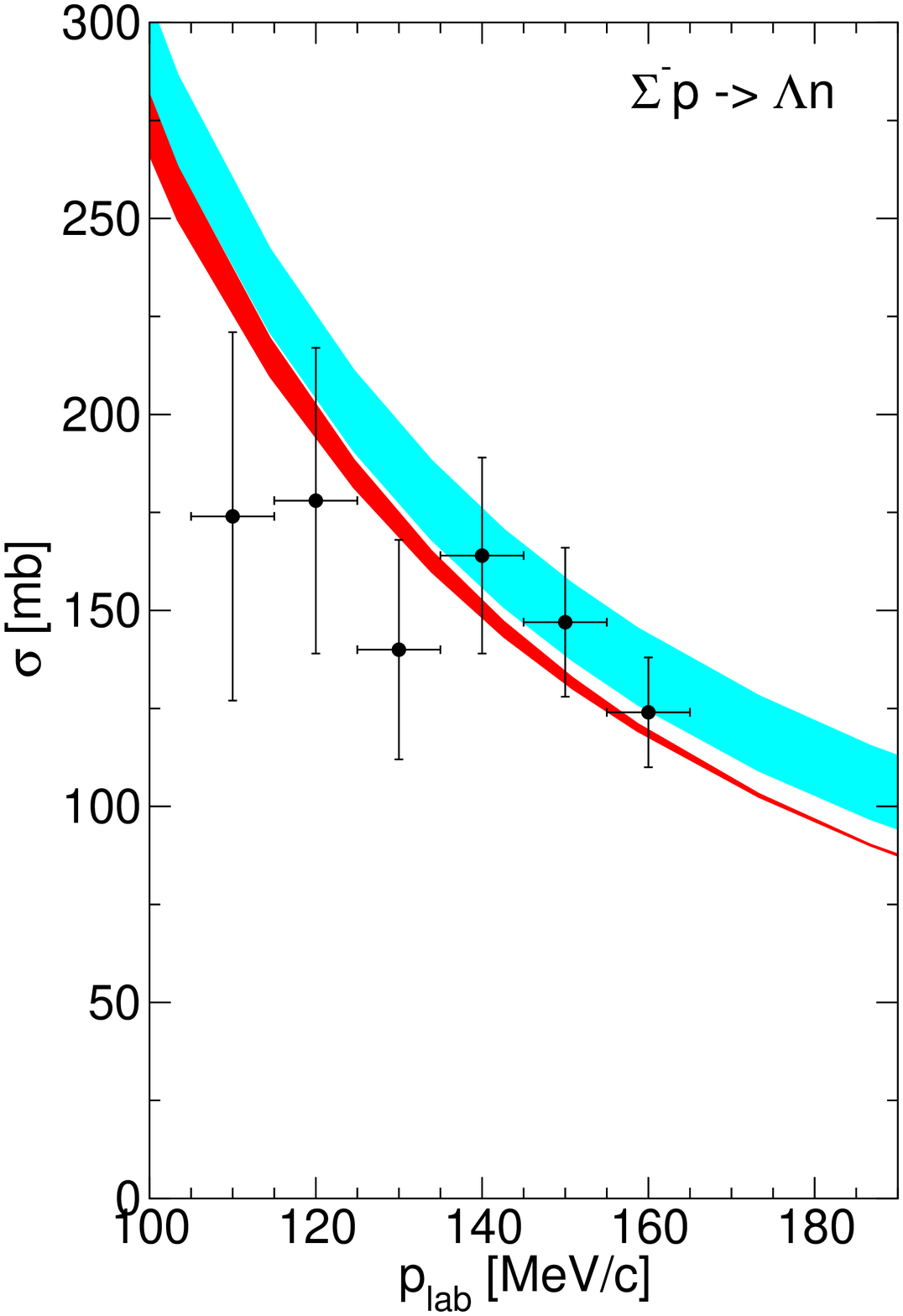}
\caption{
Total cross section $\sigma$ for $\Lambda p\to \Lambda p$ (left and middle panel)
and for $\Sigma^- p \to\Lambda n$ (right panel) as a function of $p_{{\rm lab}}$.
The red/dark band shows the chiral EFT results to NLO for variations of the cutoff
in the range $\Lambda = 500\ldots$650~MeV, 
while the cyan/light band are results to LO for $\Lambda = 550\ldots$700~MeV. 
The data can be traced back from Ref.~\cite{Haidenbauer:2013oca}.
}
\label{fig:sigYN}
\end{center}
\vspace{-5mm}

\end{figure}

In Fig.~\ref{fig:sigYN}, we show some assorted results for the reactions
$\Lambda p\to \Lambda p$ and $\Sigma^- p \to\Lambda n$. The middle panel
zooms on the region around the $\Sigma^+ n$ and the $\Sigma^0 p$ thresholds for 
$\Lambda p$ scattering. In both cases, the
bands at the given order are obtained by variations of the cut-off in the LS
equation, that is applied in terms of the regulator function $f_R(\Lambda) =
\exp[-(p'^4 + p^4)/\Lambda^4]$, with $\Lambda$ the cut-off.
This is only a very rough estimate of the theoretical uncertainty and
it should be revisited in the future upon the lines suggested e.g. in 
Refs.~\cite{Epelbaum:2014efa,Furnstahl:2015rha} However, we remark that the
NLO contributions improve the description of the data and that the uncertainty
bands shrink with increasing order.

\section{Strangeness $S=-2,-3,-4$ baryon-baryon interactions}\label{sec:YY}
\label{sec:morestr}

The experimental situation in the sector with strangeness $S=-2$ ($YY$ and $\Xi N$
interactions), with $S=-3$ ($\Xi Y$ interactions) and $S=-4$ ($\Xi \Xi$ interactions)
is even poorer than for the hyperon-nucleon interaction. There are some constraints
on the $\Lambda\Lambda$ interaction from double-$\Lambda$ hyper-nuclei and from final-state
interactions in production reactions. Similarly, there are bounds on the $\Xi^- p$ 
elastic cross section and for $\Xi^- p \to \Lambda\Lambda$ in a broad momentum range.
There are some further results on in-medium cross sections and a few constraints
from production experiments. For a detailed discussion, we refer the reader
to Ref.~\cite{Haidenbauer:2015zqb}.

 It is therefore interesting to perform exploratory studies of the baryon-baryon 
interaction with $S=-2$ and with $S=-3,-4$ at LO, 
see Refs.~\cite{Polinder:2007mp,Haidenbauer:2009qn}.
As already noted, at LO one has one more contact interaction in the $S=-2$
system compared to $S=-1$. Denoting the corresponding LEC by $C_1$ and varying
its value within the natural range, given by~\cite{Epelbaum:2005pn} $|C_1| = 4\pi /F_\pi^2$, 
with $F_\pi$ the pion decay constant, one observes that the chiral EFT predictions
are consistent with the experimental information on the $S=-2$ sector. 
{Using the LECs fixed in the fit to the $\Lambda N$ and $\Sigma N$ data,}
one  
can make predictions for the $\Xi\Lambda$, $\Xi\Sigma$ and $\Xi\Xi$ interactions.
Strong attraction is found in some of the $S=-3$ and $S=-4$ channels, suggesting the
possible existence of bound states. However, as we will argue below, these seem to 
be artefacts of the SU(3) symmetry imposed on the contact interactions.

\begin{figure}[t]
\centering
\includegraphics*[width=0.32\textwidth]{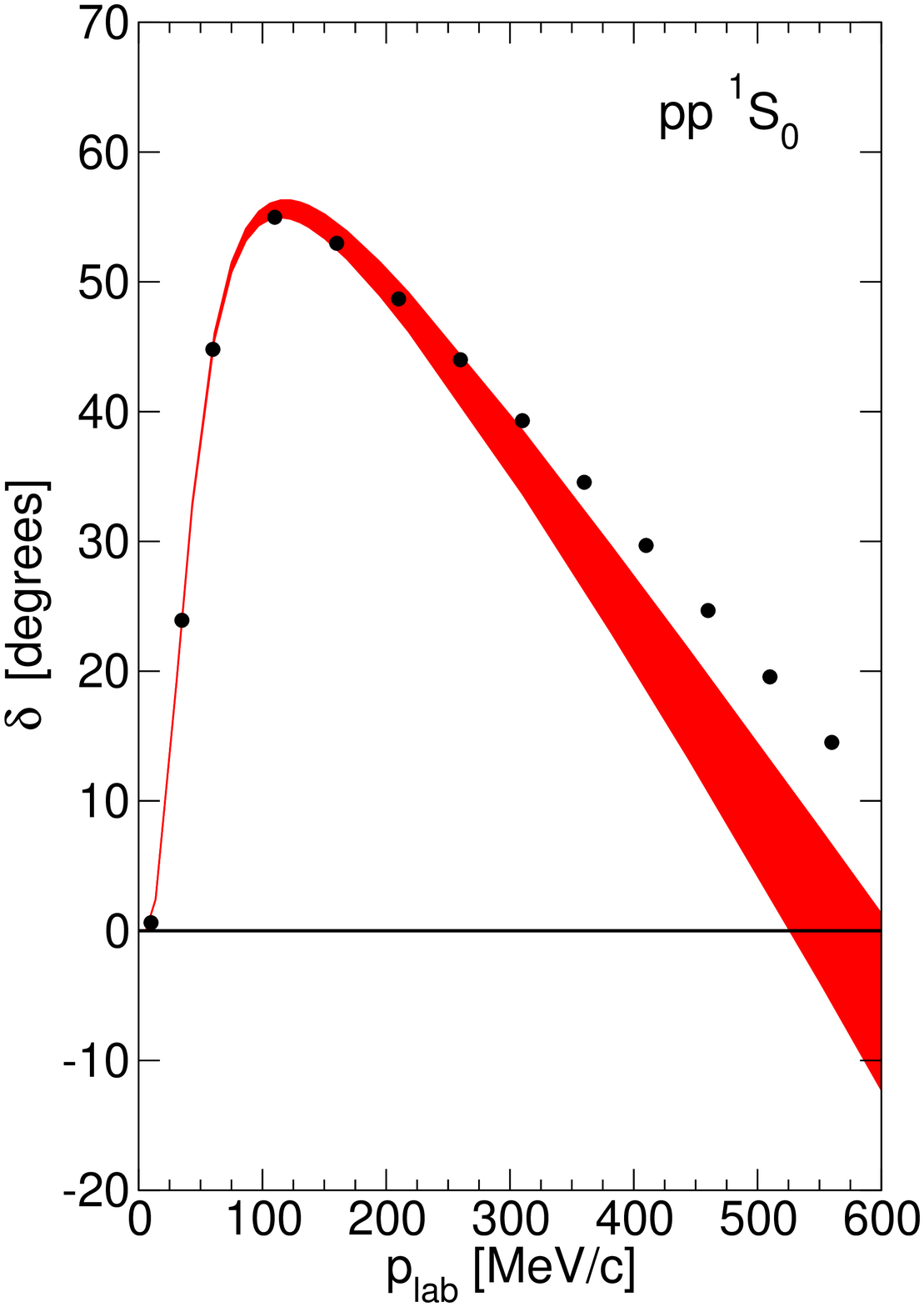}
\includegraphics*[width=0.32\textwidth]{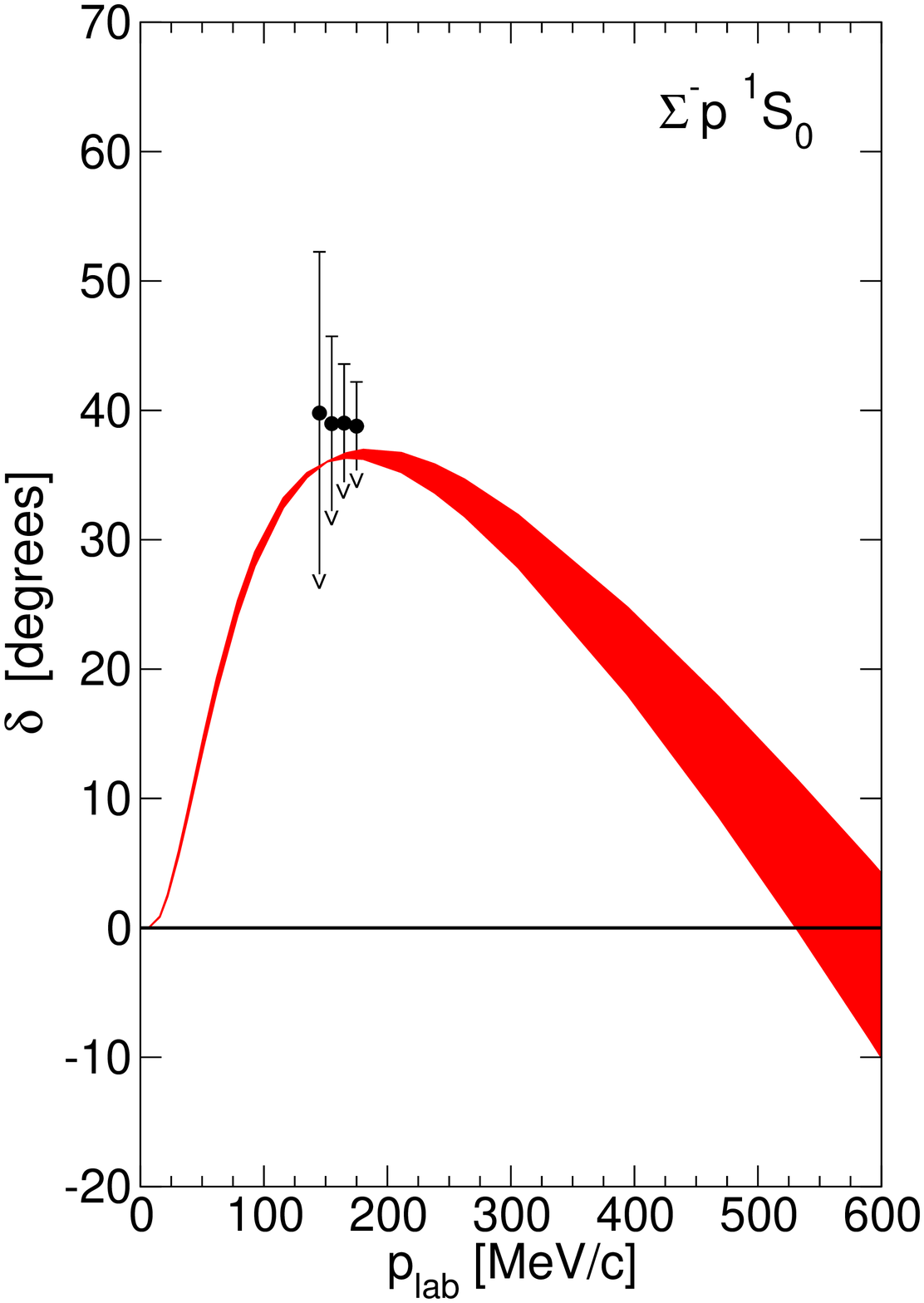}
\includegraphics*[width=0.32\textwidth]{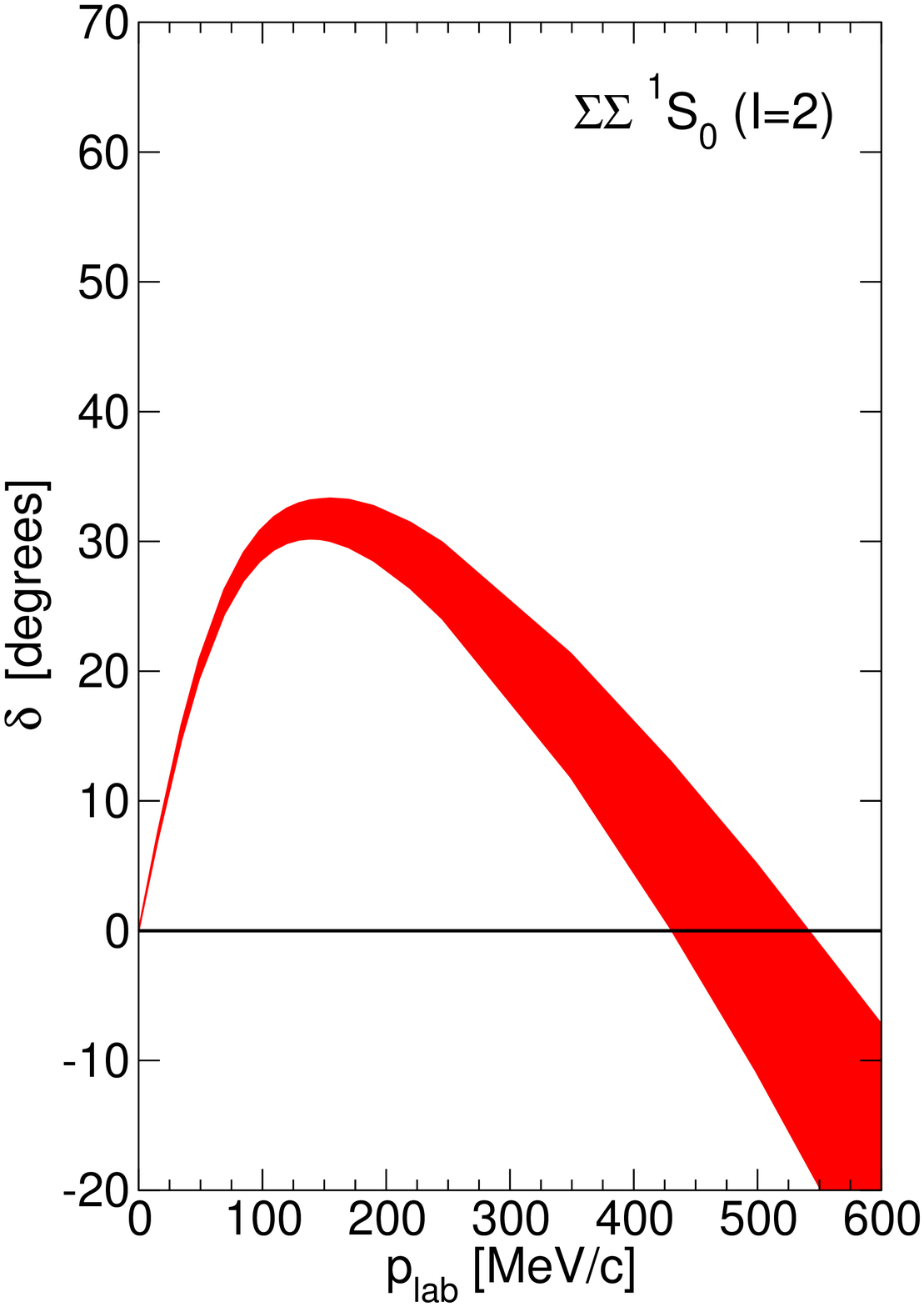}
\caption{$pp$, $\Sigma^+p$, and $\Sigma^+\Sigma^+$ phase shifts in the $^1S_0$ 
partial wave.  The filled band represent the NLO result. 
The $pp$ phase shifts of the GWU analysis \cite{SAID1} are shown by circles.
In case of $\Sigma^+ p$ the circles indicate upper limits for the phase shifts,
deduced from the $\Sigma^+ p$ cross section. 
}
\label{fig:NNphase}
\vspace{-5mm}

\end{figure}

\begin{figure}[htb]
\begin{center}
\includegraphics[width=0.35\textwidth,angle=270]{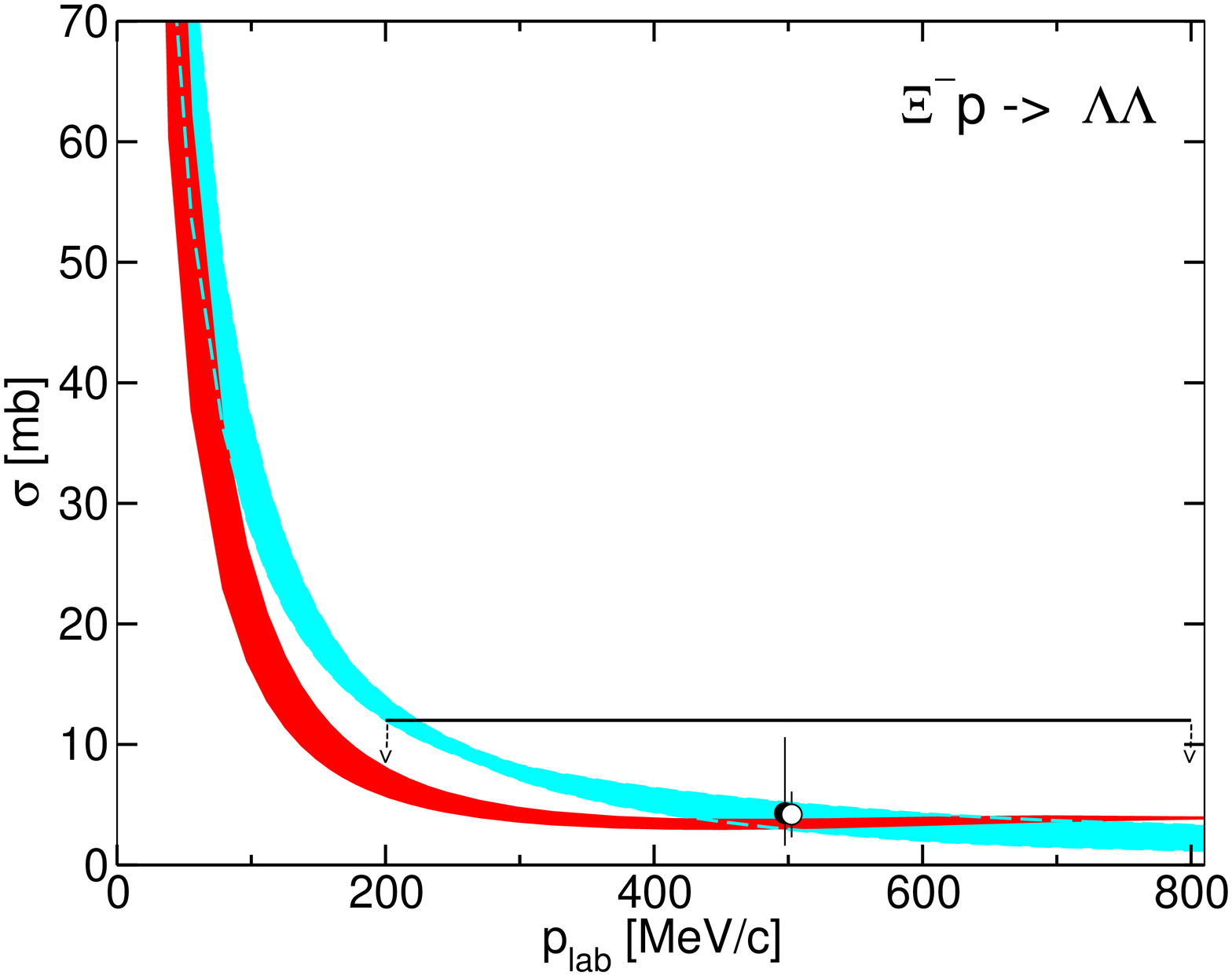}
\includegraphics[width=0.35\textwidth,angle=270]{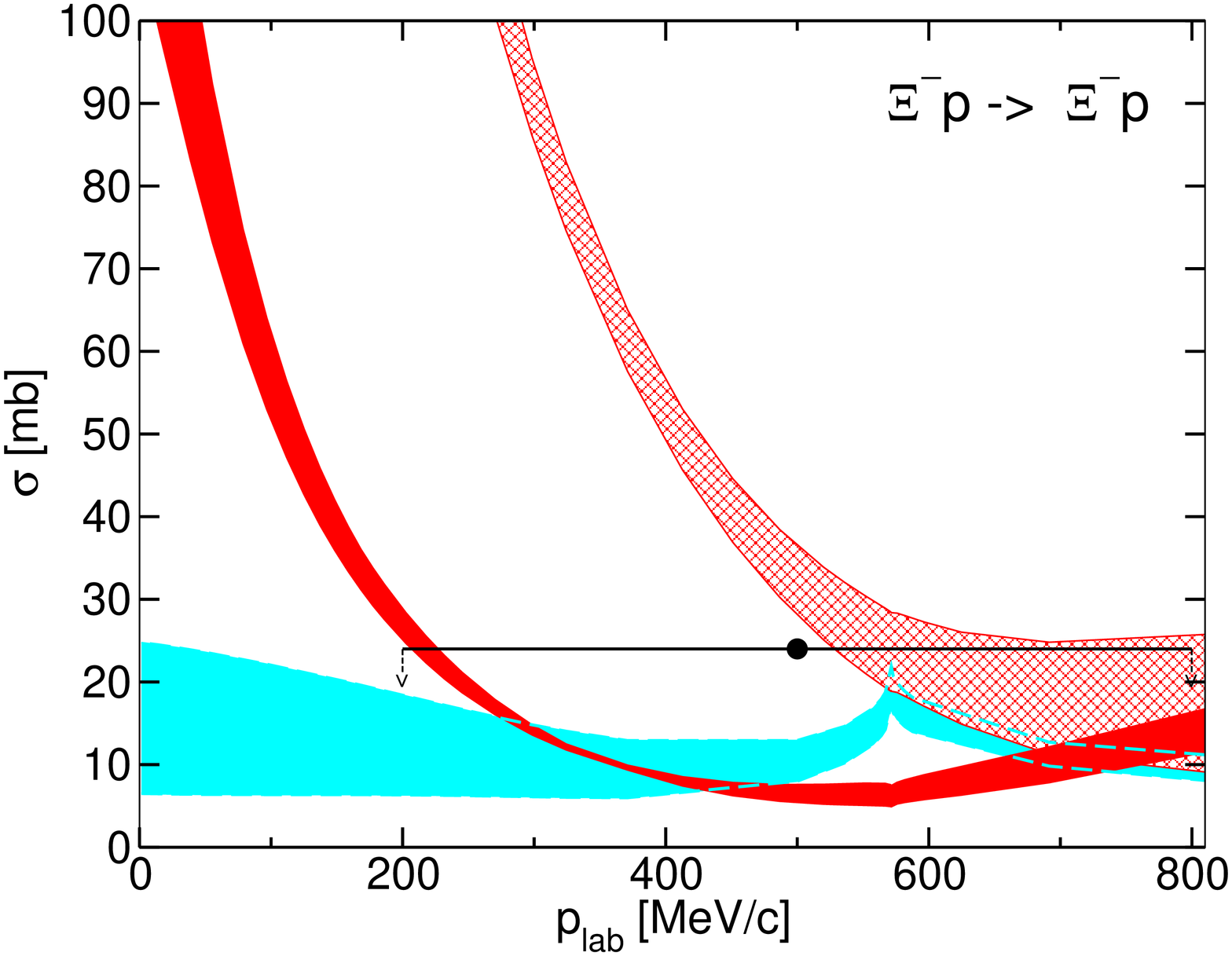}
\caption{
Total cross section $\sigma$ for $\Xi^- p\to \Lambda \Lambda$ (left panel)
and for $\Xi^- p \to\Xi^- p$ (right panel) as a function of $p_{{\rm lab}}$.
The red/dark band shows the chiral EFT results to NLO for variations of the cutoff
in the range $\Lambda = 500 \ldots$650~MeV, 
while the cyan/light band are results to LO for $\Lambda = 550 \ldots 700$~MeV. 
The hatched area in the right panel refers to a calculation where all LECs are taken
over from the $YN$ fit. The data can be traced back from Ref.~\cite{Haidenbauer:2015zqb}.
}
\label{fig:sigXN}
\end{center}
\vspace{-5mm}

\end{figure}

In the approximation of taking only the SU(3) symmetric contact interactions, a
consistent description of the $NN$ and $YN$ data is not possible. 
However, at NLO SU(3) breaking contact terms arise in the employed
Weinberg power counting scheme~\cite{Petschauer:2013uua}. These allow 
one to account for the leading {SU(3)} breaking terms in the
$^1S_0$ wave for the baryon-baryon channels with maximal isospin, that contributes
e.g. to nucleon-nucleon scattering via the potential $V_{NN}^{I=1} = C_1^\chi(M_K^2-M_\pi^2)/2$,
see Ref.~\cite{Haidenbauer:2014rna}. From a simultaneous
description of the $pp$ and the $\Sigma^+ p$ phase shifts one can pin down the
LEC $C_1^\chi$ and then predict the $\Sigma\Sigma$ interaction, see Fig.~\ref{fig:NNphase}.
In the left panel of Fig.~\ref{fig:sigXN}
we show the LO and NLO results for $\Xi^-p\to\Lambda\Lambda$ in comparison to the data
from Ahn et al.~\cite{Ahn:2006} and Kim et al.~\cite{Kim:2015S}. Both are consistent
with these data. 
Note that the LO results are solely determined by the underlying SU(3) symmetry 
whereas in the NLO case the aforementioned SU(3) symmetry breaking in the $^1S_0$ wave 
is taken into account. 
The results for $\Xi^- p\to\Xi^- p$ shown in the right panel of Fig.~\ref{fig:sigXN}  
indicate that some degree of SU(3) symmetry breaking is also required in the $^3S_1$ 
partial wave. Specifically, we see that the inclusion of the leading SU(3) symmetry
breaking term considerably lowers the prediction, giving a better agreement with the
upper bound from Ref.~\cite{Ahn:2006} (red/dark band). Relying strictly on SU(3) 
symmetry would yield cross sections that are apparently too large, cf. the hatched band. 
A more detailed account of these topics, including
also discussions of the pertinent baryon-baryon scattering lengths and effective ranges
can be found in Ref.~\cite{Haidenbauer:2015zqb}.
It is fairly obvious that to make further
progress in this field, one needs more data, that are expected to come from the J-PARC
facility in Japan and from FAIR at Darmstadt (Germany).

\section{Exotic bound states?}\label{sec:eco}

Another fascinating topic related to the baryon-baryon interactions is the possible appearance
of (exotic) bound states. 
{Historically, the so-called dibaryons have been conceived either as}
tightly bound six quark objects or as shallow bound states of two baryons. 
For the first category, the recently
observed dibaryon $d^*(2380)$ at COSY (Forschungszentrum J\"ulich)  
\cite{Adlarson:2011bh,Adlarson:2014pxj}
qualifies, whereas an established loosely bound state of a proton and the neutron, the deuteron,
is one of the best studied nuclei. The deuteron is a bound state in the $^3S_1$-$^3D_1$ channel,
whereas the binding of the proton-neutron system in the $^1S_0$ channel is just not strong enough to 
produce a bound state and only a virtual state is created.  May  be the most famous, -- 
or should we say: infamous? -- dibaryon is the
$H$-particle predicted by Jaffe within the bag model as a compact $|uuddss\rangle$ 
state~\cite{Jaffe:1976yi}. After decades of failed experiments to establish its presence, 
it regained popularity a few years ago from lattice sightings
reported in Refs.~\cite{Beane:2010hg,Inoue:2010es}, though at unphysical quark masses. 
The quantum numbers of the $H$ are these of the $\Lambda\Lambda$ system
{in an $S$-wave state}, 
namely strangeness ${S}=-2$, isospin $I=0$, 
and $J^P=0^+$.  Other predicted dibaryons involving strangeness are cousins of the 
$d^*(2380)$~\cite{Dyson:1964xwa} or generated from the attractive interaction between 
certain baryons, e.g. in the $\Xi\Xi$ but also for the 
$\Xi\Sigma$ and $\Xi\Lambda$ systems~\cite{Stoks:1999bz, Haidenbauer:2009qn}. Note that the 
NPLQCD Collaboration reported  evidence for a $\Xi^-\Xi^-$ bound  state~\cite{Beane:2011iw}.
This list is by far not exhaustive but should merely serve as an illustration that there might 
be very rich physics in systems involving strangeness - strange exotics.

The chiral EFT discussed so far is very well suited to shed light on the $H$-dibaryon, should 
it indeed exist. In particular, one can study the implications of the imposed (approximate)
SU(3) symmetry and further explore the dependence of its properties on the involved meson and
baryon masses. The latter aspect is important as the existing lattice QCD calculations were
not performed at the physical quark and thus hadron masses. In particular, one can use the
flavor singlet LEC $C_1$ as a dial to generate a bound state with a given binding 
energy~\cite{Haidenbauer:2011ah,Haidenbauer:2011za}.

\begin{figure}[t!]
\centering
\includegraphics[width=0.35\textwidth,angle=-90]{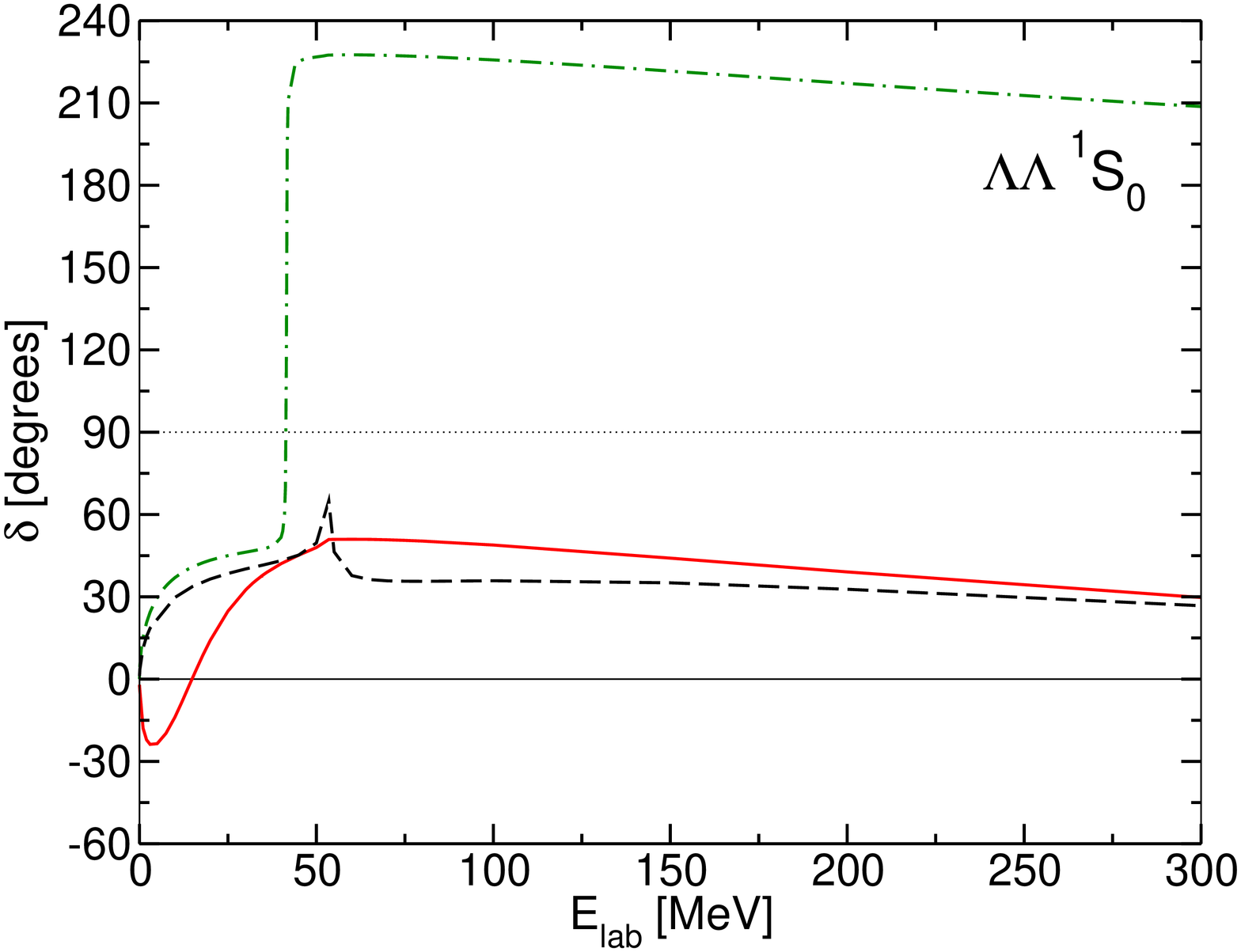}
\includegraphics[width=0.35\textwidth,angle=-90]{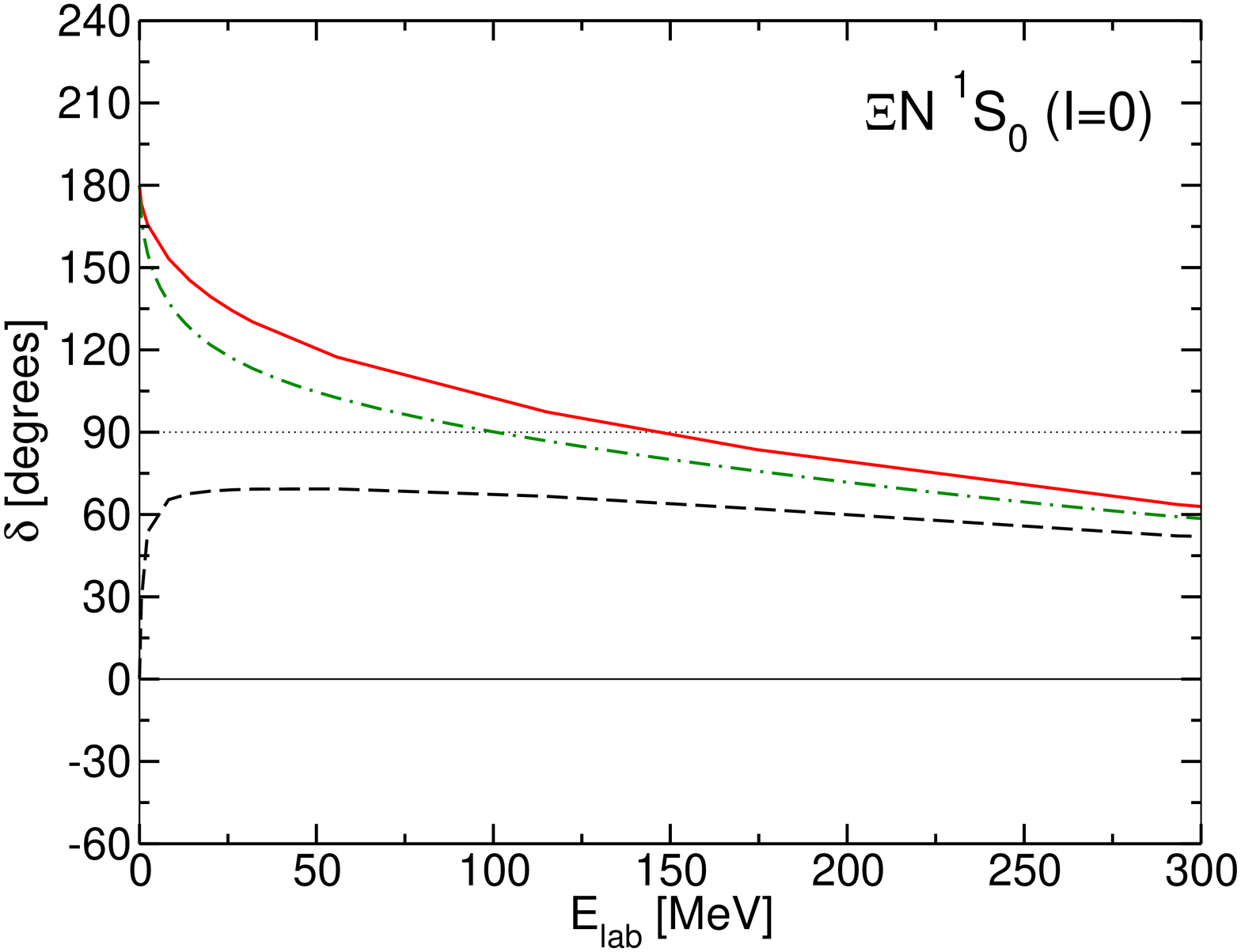}
\caption{Phase shifts for $\Lambda\Lambda$ ($^1S_0$) (left) and $\Xi N$ ($^1S_0$) (right)
as a function of the pertinent laboratory energies. 
The solid line is the result for our reference $BB$ interaction that produces a bound $H$ at
$E_H=-1.87$ MeV. The other curves are results for interactions that are fine-tuned to the 
$H$ binding energies found in the lattice QCD calculations of the HAL QCD (dashed) and 
NPLQCD (dash-dotted) collaborations, respectively, for the pertinent meson (pion) and
baryon masses as given by the lattice collaborations. 
}
\label{fig:phases}
\vspace{-2mm}

\end{figure}

The first issue to be discussed in this context is the effective range expansion in
the $^1S_0$ channel of the $\Lambda\Lambda$ interaction. Let us assume that the 
$H$ is a loosely bound two-baryon state much like the deuteron. For illustration,
let us fix the value of the LEC $C_1$ such that the binding momentum of the $H$
is the same as for the deuteron, $\gamma = 45.7\,$MeV, related to the binding energy
via $E = -\gamma^2/m_B$, with $m_B$ either $m_N$ or $m_\Lambda$. As it is well-known,
the effective range expansion of Bethe and Schwinger relates the binding momentum to
the scattering length $a$ and effective range $r$, $1/a \simeq \gamma - r\gamma^2/2$.
While this is very well fulfilled for the deuteron, the corresponding results for the
$\Lambda\Lambda$-system in the $^1S_0$ wave are very different, $a\simeq -0.65\,$fm and
$r \simeq 6\,$fm~\cite{Haidenbauer:2015zqb}. Thus, the properties of the $H$ are very different
from the ones of the deuteron, despite the fact that both are close-to-threshold bound states.
This can also be understood  from the effective potentials in the $I=0$ channel, the SU(3) 
flavor singlet $\sim C_1$ contributes with a much larger strength to $\Xi N$ than to $\Lambda\Lambda$.
This means that the $H$ should predominantly be a $\Xi N$ bound state. This can be sharpened
by looking at the corresponding phase shifts, the one in the $\Xi N$ channel is rather
similar to the $NN$ $^3S_1$ phase shift, see Refs.~\cite{Haidenbauer:2011ah,Haidenbauer:2011za},
see also Fig.~\ref{fig:phases}.

Second, one can use the chiral EFT to vary the quark/pion mass to make contact to the lattice
QCD results. If one sticks to the SU(3) symmetric case, one finds that for pion masses
below 400~MeV, the dependence of the binding energy of the $H$ on $M_\pi$ is linear, see
also Ref.~\cite{Beane:2011zpa}. In particular, if one adjusts the $H$ binding energy to 
the value found by the NPLQCD collaboration at $M_\pi=389\,$MeV, it is reduced by 7~MeV 
when going to the physical point. 
For larger pion masses, this dependence is weakened but one should be aware that the EFT
can not be trusted anymore at too large pion masses.
 
A third, and much more drastic effect, is caused by the SU(3) breaking related to the
three thresholds {for} $\Lambda\Lambda$, $\Sigma\Sigma$ and $\Xi N$,
which are located at 2231.2, 2257.7, and 
2385.0~MeV, in order. For physical values, the binding energy of the $H$ is reduced by
as much as 60~MeV as compared to an SU(3) symmetric interaction with degenerate two-baryon 
thresholds.  For the lattice QCD results of the HAL QCD collaboration, this means that the bound 
state has disappeared at the physical point where as for the NPLQCD case, a resonance 
in the $\Lambda\Lambda$ system might survive, cf. also Fig.~\ref{fig:phases}. 
The intricacies of coupled channel
systems at unphysical quark masses are further discussed in the appendix.

As noted above, there have been speculations about bound states in two-baryon systems
with $S=-3$ and $S=-4$, as e.g. given by the LO and NLO SU(3) symmetric interactions
in our EFT. However, we have rexamined the role of SU(3) breaking in Ref.~\cite{Haidenbauer:2014rna}.
As worked out by Kaiser and Petschauer~\cite{Petschauer:2013uua}, there are 12 leading order
symmetry breaking terms with corresponding LECs, from which 6 appear in the $^1S_0$ and 6 in
the $^3S_1$ partial wave. At present, there is simply not enough information to determine
these all. However, if we restrict ourselves to the $BB$ systems in the $^1S_0$ wave with maximal
isospin, only two such LECs survive, called $C_1^\chi$ and  $C_2^\chi$. We had already
discussed in Sec.~\ref{sec:morestr} how to fix $C_1^\chi$. The other LEC can not be pinned down reliably
at present, but must be varied within a reasonable range. Including the symmetry breaking
$\sim C_1^\chi$ in the $\Sigma\Sigma$ channel, the scattering length and the corresponding
attraction is much reduced so that practically any bound state in this system can be ruled out.
{It is reasonable to assume}  
that $C_2^\chi$ is of the same size as $C_1^\chi$ and that the increase in repulsion
when going from $NN$ to $\Sigma N$ to $\Sigma\Sigma$ is not reversed for the $S=-3$ and $S=-4$ 
systems. In that case, bound states in these systems are also rather unlikey. For a more
detailed discussion, we refer the reader to Ref.~\cite{Haidenbauer:2014rna}.

\section{Hyperons in nuclear matter}\label{sec:med}

 Apart from the quest to find repulsion in the $YN$ interaction, there are other interesting aspects
 of the behavior of hyperons in nuclei or nuclear matter. In particular, the repulsive nature of
 the $\Sigma$-nucleus potential \cite{Friedman07}  and the weak $\Lambda$-nucleus spin-orbit interaction 
 \cite{HT06,Gal10,Botta} are long-standing issues, on which our calculations of hyperons in nuclear 
 matter have shed some light~\cite{Haidenbauer:2014uua,Petschauer:2015nea}.
 
 First, we consider the antisymmetric spin-orbit force. It is generated from the NLO potential
 $V \sim (\vec\sigma_1 - \vec\sigma_2)\cdot(\vec q \times \vec k)$, in terms of the Pauli
 spin matrices, the exchange momentum $\vec q$ and the total momentum $\vec k$.
 This term gives rise to (spin) singlet-triplet ($^1P_1$-$^3P_1$) transitions. 
 {In our study of the $YN$ interaction in free space~\cite{Haidenbauer:2013oca}} 
 this antisymmetric spin-orbit force was set to zero because it can not be determined from $YN$ 
 scattering data. 
 Matters are different in nuclei. Here, the {strength of the} spin-orbit interaction is frequently
 parameterized in terms of the so-called Scheerbaum factor $S_Y$~\cite{SCHE}, which is related to
 the {hyper-nuclear spin-orbit}  
 potential via $U_Y^{\ell s}(r) = - (\pi/2) S_Y (1/r) (d\rho(r)/dr) \vec\ell\cdot\vec\sigma$,
 with $\rho(r)$ the nucleon density distribution and $\vec\ell$ the single-particle orbital 
 angular momentum operator. 
 The EFT approach allows us to tune the LEC related to the $^1P_1$-$^3P_1$ transition 
to achieve a value of~\cite{Haidenbauer:2014uua} $S_\Lambda = -3.7\,$MeV~fm$^5$,
 in accordance with phenomenological determinations that give $S_\Lambda$ in the range $-4.6$ to
 $-3.0\,$MeV~fm$^5$.~\cite{Kohno:2009sc,Kohnopc}
 
 The standard method to calculate the properties of hyperons in a nuclear medium is Brueckner
 theory. The Brueckner reaction matrix ($G$-matrix) is determined from a solution of the
 Bethe-Goldstone equation, $G(\omega) = V + V[Q/(e(\omega) + i\epsilon)] G(\omega)$, with
 $V$ the pertinent free-space potential, $e(\omega)$ the energy denominator depending on the
 starting energy $\omega$, and $Q$ is the (angle-averaged) Pauli operator. Medium effects are
 thus generated from the Pauli operator as well as the density-dependent single-particle potential
 $U(\omega)$ in the energy denominator $e(\omega)$. This single-particle potential is 
 obtained self-consistently from the $G$-matrix. There are two commonly used methods to calculate  the
 $G$-matrix. In the so-called gap choice, only the free particle energies of the intermediate states
 appear in the energy denominator of the Bethe-Goldstone equation. This method was e.g. used 
 in Ref.~\cite{Haidenbauer:2014uua}. In the so-called continuous choice, the dependence of the energy 
 denominator on the full single-particle energies is retained. This method is computationally more difficult but
 allows to reliably access the imaginary parts of the single-particle potentials, which is not possible 
 when the gap choice is employed. The results presented in Ref.~\cite{Petschauer:2015nea} are based 
 on the continuous choice.
 
\begin{figure}[t!]
\centering
\includegraphics[width=0.35\textwidth,angle=-90]{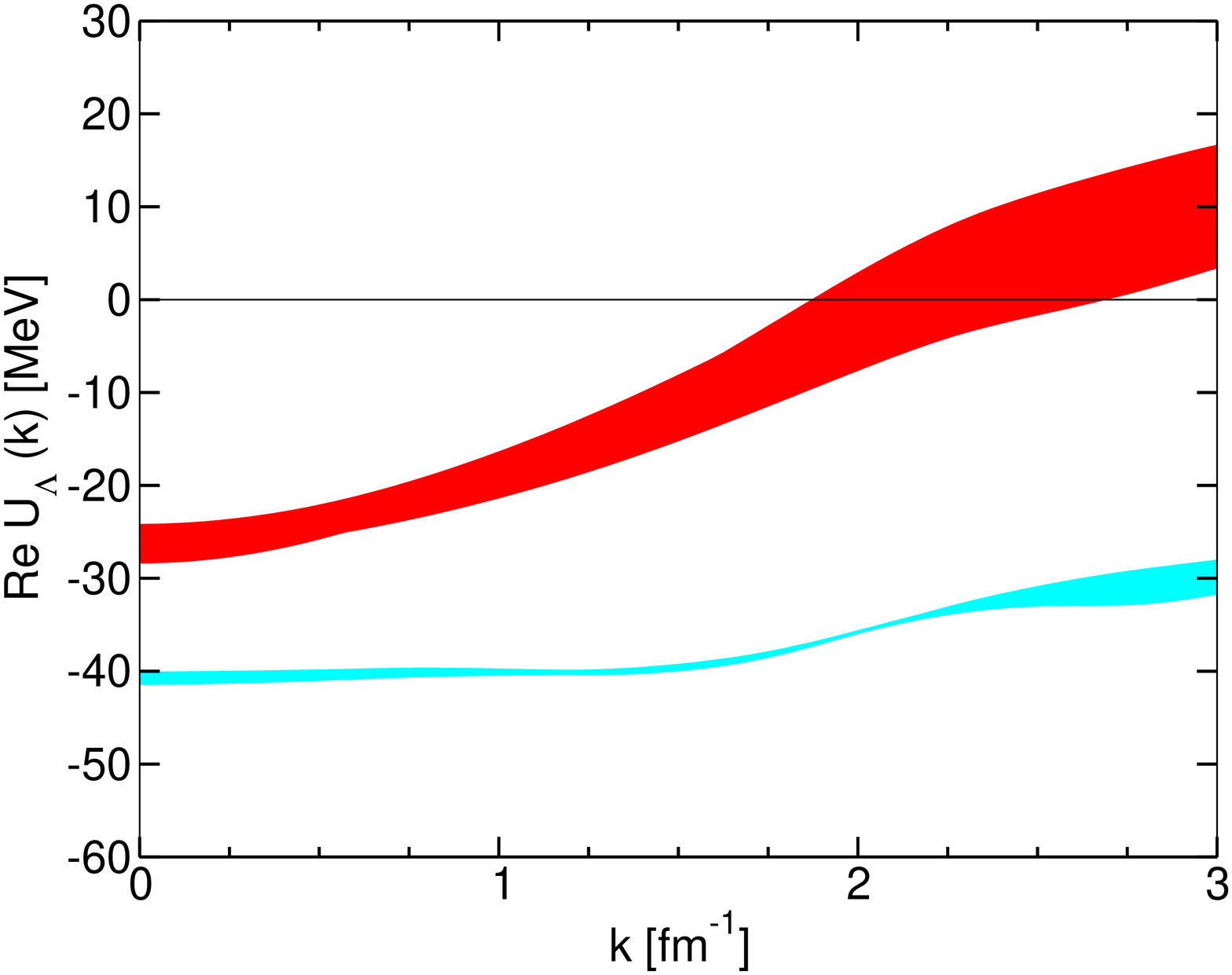}
\includegraphics[width=0.35\textwidth,angle=-90]{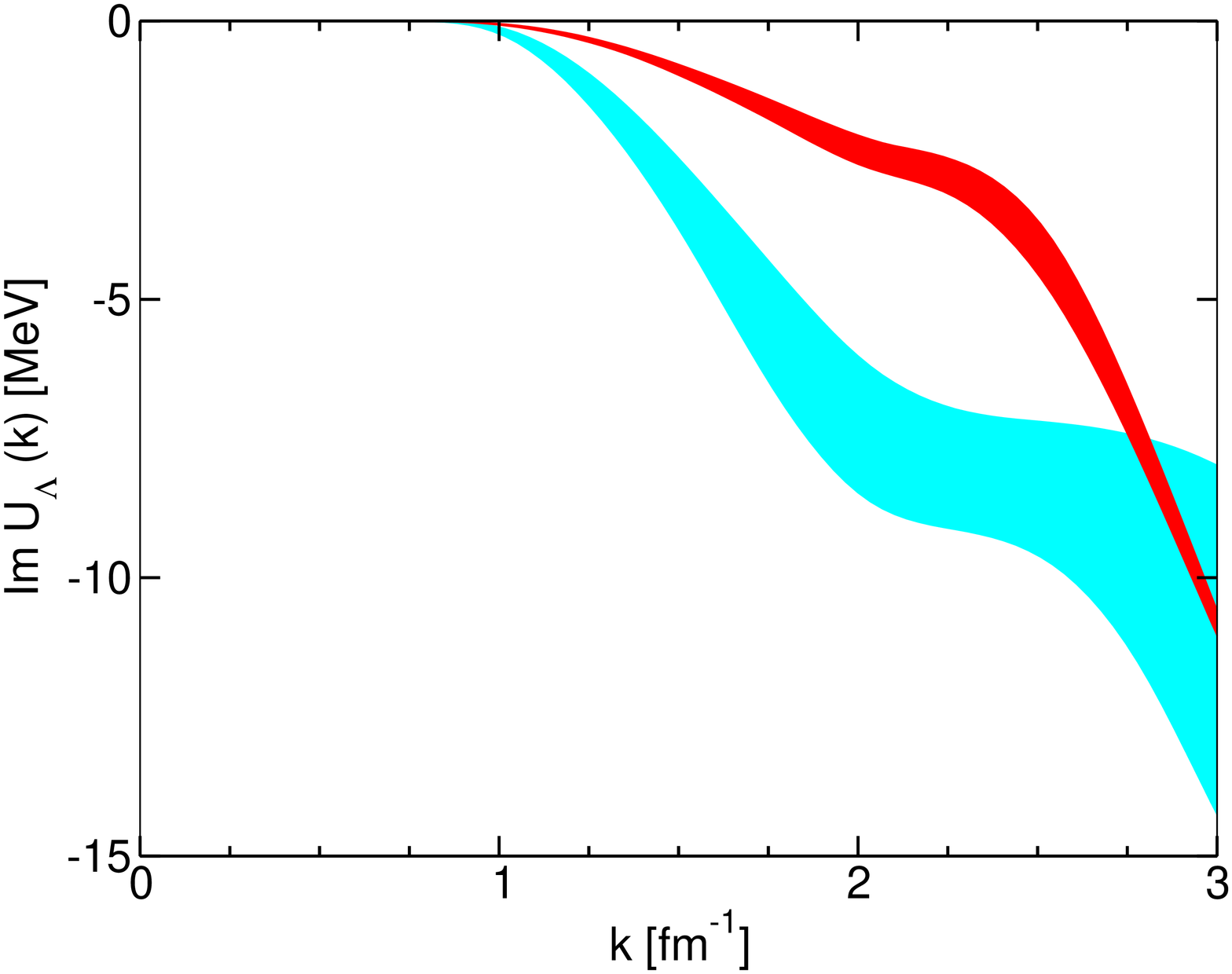}
\caption{Momentum dependence of the real and imaginary parts of the single-particle 
potential of a $\Lambda$ hyperon in isospin symmetric nuclear matter at saturation density.
The bands represent the variation of our results with the cutoff, see text.}
\label{fig:UL}
\vspace{-2mm}

\end{figure}

\begin{figure}[ht]
\centering
\includegraphics[width=0.35\textwidth,angle=-90]{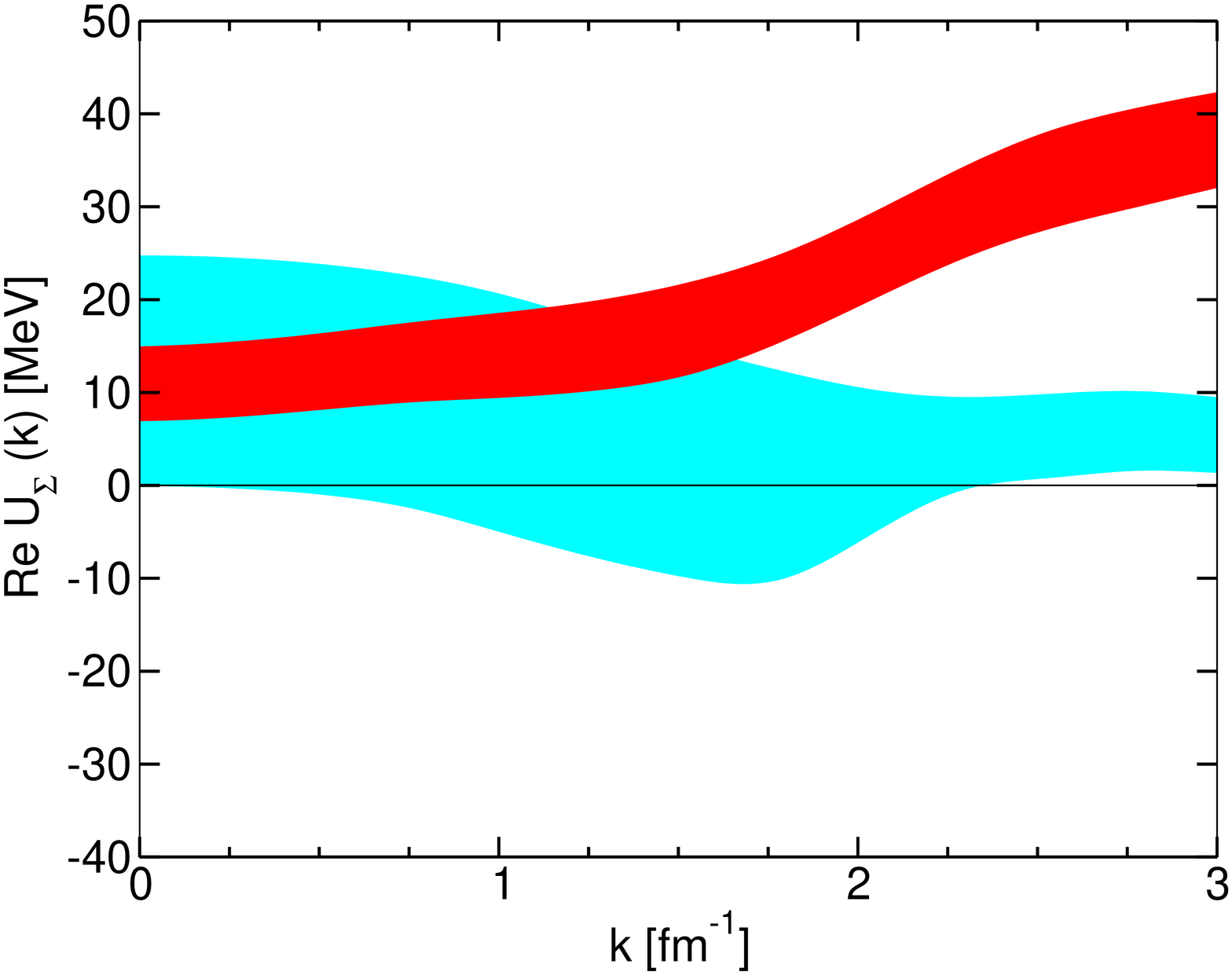}
\includegraphics[width=0.35\textwidth,angle=-90]{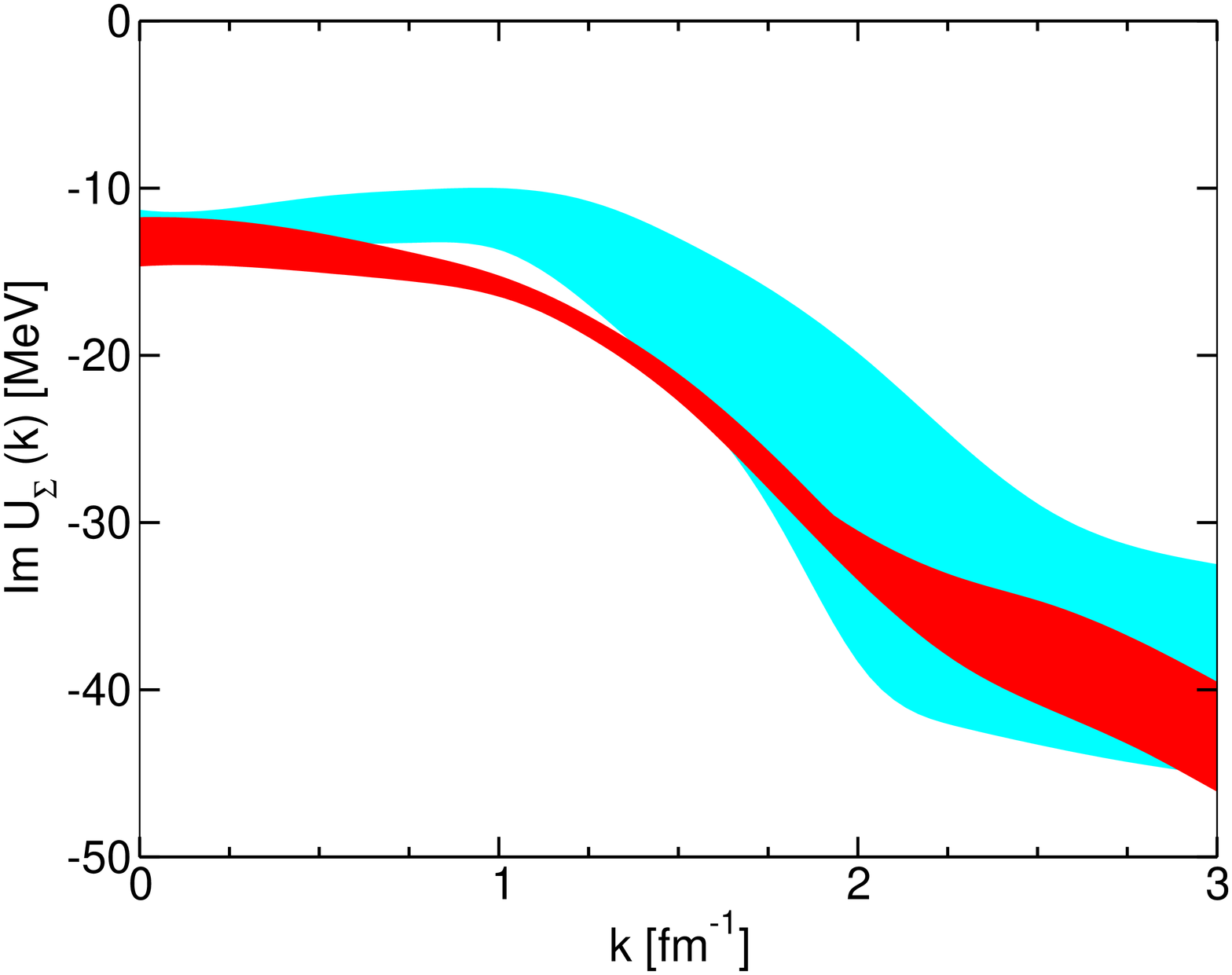}
\caption{Momentum dependence of the real and imaginary parts of the 
single-particle potential of a $\Sigma^0$ hyperon in isospin symmetric nuclear 
matter at saturation density. The bands represent the variation of our results 
with the cutoff, see text.}
\label{fig:US}

\end{figure}

 In Fig.~\ref{fig:UL}, we show the momentum dependence of the  real and the imaginary part of the
 $\Lambda$ single-particle potential in isospin-symmetric nuclear matter. The bands refer to the
 usual cut-off variations, using the continuous choice. We find $U_\Lambda (k=0) = -28 \ldots -24\,$MeV,
 consistent with the empirical value of about $-28\,$MeV as deduced from binding energies of $\Lambda$ 
 hyper-nuclei\cite{Millener1988,Yamamoto1988}. 
 The corresponding results for the neutral $\Sigma$ in isospin-symmetric matter at saturation density 
 are shown in Fig.~\ref{fig:US}, with small differences
 for the charged $\Sigma$ hyperons given by the small inter-multiplet mass splittings. In pure neutron matter,
 matters are very different. Due to the maximal asymmetry between protons and neutrons, the single-particle
 potentials for the $\Sigma^+, \Sigma^0, \Sigma^-$ hyperons are rather different. These are discussed in 
 detail in Ref.~\cite{Petschauer:2015nea}. 
{As already said above, the} resulting {$\Lambda$-nuclear} spin-orbit potential is small, which is
 partly due to cancellations between contributions from the symmetric and the anti-symmetric spin orbit forces,
 and partly due to the repulsive interactions in some of the $P$-waves. 
 The corresponding Scheerbaum factors {for the $\Sigma$ hyperons} are much bigger than
 for the $\Lambda$. {Specifically,} for the neutral $\Sigma$ we find $S_\Sigma = -15 \ldots -18\,$MeV~fm$^5$ at NLO. 
 For more details, especially also a comparison between the results obtained in the gap and the
 continuous choice, see Refs.\cite{Haidenbauer:2014uua,Petschauer:2015nea}.  
{Note that the in-medium} hyperon spin-orbit interaction was also
 investigated in Refs.~\cite{Kaiser:2008gq,Kohno:2009sc,Ishii14ls}. 

\section{Three-baryon interactions}\label{sec:3bf}

The need for (repulsive) three-baryon forces was already alluded to in the introduction.
Another indication that points towards the necessity of such forces 
are the calculations of the binding energies of light hyper-nuclei
at NLO by Nogga~\cite{Nogga:2014kma}. Here, the remaining cut-off dependence hints at
missing three-body forces. More generally, it is well established that three-body
forces are required in nuclei and nuclear matter, see e.g. 
Refs.~\cite{KalantarNayestanaki:2011wz,Hammer:2012id}.

\begin{figure}[t!]
\centering
\includegraphics[width=0.725\textwidth]{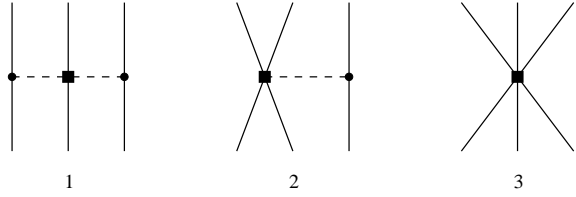}
\vspace{-2mm}
\caption{Toplogies of the leading three-baryon forces. (1) Two-pion exchange terms, (2)
one-pion exchange terms, and (3) six-baryon contact terms.}
\label{fig:3nf}
\vspace{-2mm}

\end{figure}

In the Weinberg power counting, three-baryon forces (3BFs) appear at next-to-next-to-leading
order and are given in terms of the three topologies shown in Fig.~\ref{fig:3nf}. The first
and second topology require the meson-baryon Lagrangian at NLO. Its minimal form is given
in Ref.~\cite{Frink:2006hx} (see also Refs.~\cite{Krause:1990xc,Oller:2007qd}).
The six-baryon contact terms given by the topology (3)
require the general SU(3) Lagrangian that was constructed by 
Kaiser and Petschauer~\cite{Petschauer:2013uua}. Armed with these ingredients, the leading
three-baryon forces were derived in Ref.~\cite{Petschauer:2015elq} within SU(3) chiral EFT.
For that, the chiral Lagrangian in the non-relativistic limit with the minimal  number
of terms in the full SU(3) sector was derived. One finds that there are in total 18 different
structures corresponding to topology~(3) accompanied by 18 LECs. However, only 
a limited number of combinations of these LECs are involved in a given process. 
In general, one can write this local contribution to the 3BFs as 
\begin{equation}
V^{\rm ct} = N_1 \mathbb{1} + N_2 \vec\sigma_1\cdot\vec\sigma_2 + N_3 \vec\sigma_1\cdot\vec\sigma_3 
         + N_4 \vec\sigma_2\cdot\vec\sigma_3 + N_5 i \vec\sigma_1\times \vec\sigma_2\cdot\vec\sigma_3~,
\end{equation}
in terms of the spin matrices of the three baryons and the $N_i$ are prefactors containing LECs and
kinematical factors, and we have suppressed the corresponding isospin factors, as detailed in
Ref.~\cite{Petschauer:2015elq}. 
It can be shown that in the SU(2) limit one recovers the purely
nucleonic contact term that is accompanied by the LEC $E$.\cite{Epelbaum:2002vt}
The graphs corresponding to the one-meson exchange topology (2) require the knowledge
of combinations of the 14 four-baryon-one-meson vertices listed in Ref.~\cite{Petschauer:2013uua}.
In spin-momentum space, this translates into the following structures:
\begin{equation}
V^{\rm 1\phi} = -\frac{1}{2F_\pi^2}\frac{\vec\sigma_1\cdot\vec{q}_1}{\vec{q}_1^{\, 2}+M_\phi^2}
  \Bigl\{ N_6 \vec\sigma_2\cdot\vec{q}_1 + N_7 \vec\sigma_3\cdot\vec{q}_1 + N_8
    (\vec\sigma_2\times \vec\sigma_3)\cdot\vec{q}_1\Bigr\}~,      
\end{equation}
where the $N_i$ are again composed of LECs and kinematical factors. 
Further, $\vec{q}_i = \vec{p}_i^{~\prime}-\vec{p}_i$, with $p_i$ and $p_i'$ the
initial and final momenta of baryon $i$, respectively.
Again, in the SU(2)
limit, one recovers the well-known term that is parameterized in terms of the LEC $D$,
that can e.g. be determined from single pion production~\cite{Hanhart:2000gp}.
Finally, the diagrams corresponding to topology~(1) are the generalizations of the
well-known Fujita-Miyazawa three-nucleon force, when the decuplet is considered as
very heavy. The spin-momentum structure of these terms takes the form
\begin{eqnarray}
V^{\rm 2\phi} &=& -\frac{1}{4F_\pi^2}\frac{\vec\sigma_1\cdot\vec{q}_1~~ \vec\sigma_3\cdot\vec{q}_3}
             {(\vec{q}_1^{\, 2}+M_{\phi_1}^2)(\vec{q}_3^{\, 2}+M_{\phi_3}^2)}\nonumber\\
           &\times& \Bigl\{N_9 M_\pi^2 + N_{10}M_K^2 + N_{11} \vec{q}_1\cdot\vec{q}_3
                           + N_{12}  \vec\sigma_2\cdot ( \vec{q}_1\times\vec{q}_3)\Bigr\}~, 
\end{eqnarray}
in the same notations as before. 
Again, in the SU(2) limit this expression recovers the well-known
result in terms of the pion-nucleon LECs $c_{1,3,4}$, cf. e.g. Refs.~\cite{Friar:1998zt,Epelbaum:2002vt}.
The complete expressions for the much thought after $\Lambda NN$ three-body force can be found
in Ref.~\cite{Petschauer:2015elq}.

\begin{figure}[tb]
\centering
  \includegraphics[width=.705\textwidth]{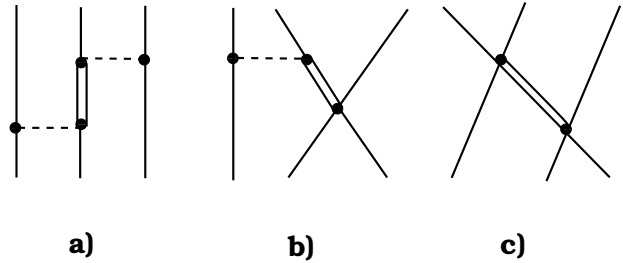}~~~
  \caption{
  Leading contributions to the three-baryon force at NLO in the theory with decuplet
   baryons. Solid, double and dashed lines denote octet baryons, 
           decuplet baryons and Goldstone bosons, in order. The circles represent
           leading vertices. The diagram of type b) with a decuplet baryon 
           in the initial state is not shown.
}
\label{fig:delta}
\vspace{-3mm}

\end{figure}

At first sight, the large number of LECs (or combinations thereof) seems to be
discomforting. It certainly is not possible to fix all these from data at present.
However, one can use decuplet saturation for estimating these LECs. This is
based on the successful resonance saturation hypothesis in the pion-nucleon
sector~\cite{Bernard:1996gq} that has already been utilized to estimate 
nucleonic three-body forces in Refs.~\cite{vanKolck:1994yi,Friar:1998zt,Epelbaum:2007sq}.
For that, consider the theory with explicit decuplet degrees of freedom. Here, the three-body
forces appear already at NLO, and the pertinent diagrams are shown in Fig.~\ref{fig:delta}.
{These diagrams involve two kinds of vertices. First, there is the leading 
meson-baryon-baryon vertex with an octet baryon and a decuplet baryon (graphs~(a) and~(b)). 
The corresponding LEC ${\mathcal C}$ (sometimes also called $h_A$) can be estimated from 
the decay width $\Delta \to N \pi$, and one finds ${\mathcal C} = 3g_A/4 \simeq 1$, with $g_A$ 
the nucleon axial-vector coupling constant. SU(3) symmetry can then be used to fix the
values for all other possible combinations of octet and decuplet baryons.  
The other kind of vertices are the ones with three octet baryons and one decuplet 
baryon (appearing in graphs~(b) and~(c)). Those vertices involve two new constants. 
In this case the pertinent constants can not be deduced from considering the $3N$-$\Delta$ 
vertex simply because in leading order the latter is Pauli forbidden.
}
That there {are} exactly two such terms can be easily understood from group
theoretical considerations, for details see Ref.~\cite{Pet} Let us call the corresponding LECs
$G_1$ and $G_2$ and consider the various topologies in detail and their behaviour
in the limit of infinite decuplet masses, 
keeping the ratio of coupling constants over decuplet mass fixed. 
The graph~(a) is nothing but the dominant contribution to the famous Fujita-Miyazawa
force. Its emergence in chiral EFT is discussed in detail in Ref.~\cite{Meissner:2008zza} In the
heavy decuplet limit, this type of graphs generates the topology~(1) of Fig.~\ref{fig:3nf}, with the
LECs in the theory without decuplet given in very symbolic notation by $C^{\rm 2\phi} \sim
\alpha_{\rm 2\phi} {\mathcal C}^2/\Delta$,
where $\alpha_{\rm 2\phi}$ is some numerical factor and $\Delta$ denotes the decuplet-octet 
mass splitting. Similarly,
the one-pion exchange topology~(2) in the theory without the decuplet is generated from the terms
of the type~(b) in Fig.~\ref{fig:delta}, leading to LECs with their strengths given 
by $C^{\rm 1\phi} \sim \alpha_{\rm 1\phi}{\mathcal C} (G_1 + \beta_{\rm 1\phi} G_2)/\Delta$, 
again with $\alpha_{\rm 1\phi}$ and $\beta_{\rm 1\phi}$ numerical factors. Finally, the graphs
 of the type~(c) lead to local six-baryon contact interactions with their LECs given by
$C^{\rm ct} \sim \alpha_{\rm ct} (G_1+ \beta_{\rm ct} G_2)^2/\Delta$. Explicit expressions
for the $\Lambda NN$ force can be found in  Ref.~\cite{Pet}.  Eventually, these LECs might also be
computed directly from lattice QCD, say along the lines of Ref.~\cite{Beane:2007es}. We are looking 
forward to such calculations. To end this section, we stress that it will be of utmost 
interest to work out the consequences of these three-baryon forces within hyper-nuclei 
and compact dense objects like neutron stars.

\section*{Acknowledgements}

We thank Tom Kuo and Ismail Zahed for inviting us to write this paper.
We acknowledge fruitful collaborations with Norbert Kaiser, Andreas Nogga,
Stefan Petschauer and Wolfram Weise.
This work was supported in part by DFG and NSFC (CRC~110), by the
HGF Virtual Institute NAVI (grant no. VH-VI-417)  and 
by the Chinese Academy of Sciences (CAS) President's
International Fellowship Initiative (PIFI) (Grant No. 2015VMA076).

\appendix
\section{Coupled channel dynamics on the lattice: Intricacies}

Here, we briefly summarize the work of Ref.~\cite{Doring:2013glu} that
nicely exhibits the intricacies one faces when one is dealing with a
coupled channel system at unphysical quark masses in a finite volume.
The starting point is the observation, first made by the Munich group,~\cite{Kaiser:1995cy}
that the baryon resonance $S_{11}(1535)$ can be generated through coupled
channel dynamics in the  $\pi N$, $\eta N$, $K\Lambda$ and $K\Sigma$
systems with total isopin $I=1/2$ in the odd-parity 
 $S_{11}$ partial wave of pion-nucleon scattering.
Including also NLO terms in the interaction kernel and using a field theoretical
regularization method, this calculation was sharpened in Ref.~\cite{Bruns:2010sv},
where it was shown that also the next resonance, the  $S_{11}(1650)$, is generated
dynamically, see also Ref.~\cite{Nieves:2001wt}  for a similar conclusion.

\begin{figure}[t!]
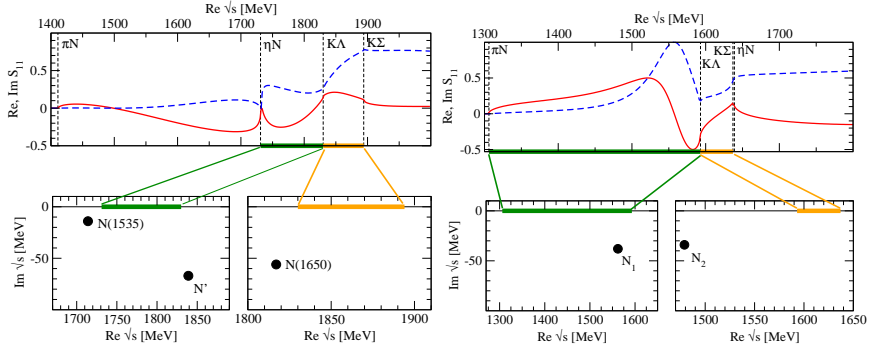

 \includegraphics[width=0.49\linewidth]{etmc.eps}
 \includegraphics[width=0.49\linewidth]{qcdsf.eps}
 \caption{Left upper panel: Real (solid line) and imaginary part (dashed line) of the $S_{11}$ amplitude, chirally extrapolated using masses and decay
 constants of the ETM collaboration. Left lower panels: 
two of the Riemann sheets with poles (left:  Riemann sheet connected to 
the physical axis between the $\eta N$ and the $K\Lambda$ threshold. right:
sheet connected to the physical axis between $K\Lambda$ and $K\Sigma$ threshold.).
Right upper panel: Real (solid line) and imaginary part (dashed line) of 
the $S_{11}$ amplitude, chirally extrapolated using masses and
decay constants of the QCDSF collaboration~\cite{Bietenholz:2011qq}. 
Left lower panels: two of the Riemann sheets with poles (same as for ETMC).
}
 \label{fig:etmc}
\vspace{-3mm}

\end{figure}

In Ref.~\cite{Doring:2013glu} we have considered two lattice set-ups that allow
to investigate the rich phenomenology in the odd-parity $S_{11}$ partial wave for
varying quark masses. Set~A is related to the work of the European Twisted 
Mass collaboration (ETMC). 
In this set-up the meson masses and pion decay constant are taken from the recent 
calculation in $N_f=2+1+1$ twisted mass lattice QCD, i.e.
ensemble $B25.32$ of Ref.~\cite{Ottnad:2012fv}. For the lattice size of $L/a=32$ and 
spacing $a=0.078$~fm, the pion mass is fixed there to
$M_\pi=269$~MeV, whereas the strange quark mass is held approximately at the physical 
value. As the kaon and eta decay constants are not
available in this calculation at the moment, we decided to relate them to $F_\pi$ with 
typical ratios of $1.15$ and $1.3$, respectively. The
baryon masses are also taken from a calculation by the ETMC, 
however, with only two dynamical quarks and an older lattice action,
see Ref.~\cite{Alexandrou:2009qu}. Nevertheless, the strange quark mass is 
held again approximately at the physical value and $M_\pi=269$~MeV
for the identical lattice size and comparable lattice spacing, i.e. $a=0.0855$~fm.
The $S_{11}$ amplitude, with the masses and decay constants of the ETMC, 
is shown in the left upper panel of Fig.~\ref{fig:etmc}.
Comparing to the physical situation, all thresholds have moved to higher energies. 
The cusp at the $\eta N$ threshold has become more pronounced, but
no clear resonance shapes are visible. The structure of the amplitude becomes clearer 
by inspecting the complex energy plane on different
Riemann sheets. This is visualized in the lower left panels of Fig.~\ref{fig:etmc}.
Compared to the physical point, the imaginary parts of the pole positions became much smaller due to the reduced phase space. Both the
thresholds and the real parts of the pole positions have moved to higher energies. However, the thresholds have moved farther than the pole
positions, such that the $N(1535)$ and $N(1650)$ poles are no longer situated below the part of the respective sheet, that is connected to the
physical axis (thick horizontal lines). The poles are thus hidden and no clear resonance signals are visible in the physical amplitude.
Instead, the amplitude is dominated by cusp effects.

Matters are very different for set~B, that refers to calculation from 
QCDSF~\cite{Bietenholz:2011qq}.  Here, baryon and meson masses
are determined from an alternative approach to tune the quark masses,
namely to start with the SU(3) symmetry limit and work at a fixed sum of the
quark masses. Most  importantly, while the lattice size and spacing are comparable to
those of the ETMC, i.e. $L/a=32$ and $a=0.075$~fm, the strange quark mass differs 
significantly from the physical value. The latter results in a
different ordering of the masses of the ground-state octet mesons and, 
consequently, in a different ordering of meson-baryon thresholds. For
further details we refer the interested reader to Ref.~\cite{Bietenholz:2011qq}.
The amplitude using the QCDSF parameter set is shown in the left upper panel of 
Fig.~\ref{fig:etmc}. In contrast to the ETMC case, a clear resonance
signal is visible below the $K\Lambda$ threshold, that is the first 
inelastic channel in this parameter setup. Indeed, we find a pole $N_1$ on
the corresponding Riemann sheet, as indicated in the right lower first panel. 
Unlike in the ETMC case, it is not hidden behind a threshold. Between
the $K\Lambda$ and the $K\Sigma$ threshold, there is only the hidden pole $N_2$ 
(right lower second panel). The $K\Sigma$ and $\eta N$ thresholds are almost
degenerate and on sheets corresponding to these higher-lying thresholds we 
only find hidden poles.

This shows that the extraction of the scattering amplitude from lattice QCD data
is a major challenge as we demonstrate by extrapolating the physical
$S_{11}$ amplitude of pion-nucleon scattering
to the finite volume and unphysical quark
masses, using a unitarized chiral framework including all next-to-leading
order contact terms. As shown, the pole movement of the resonances
$N(1535)1/2^-$ and $N(1650)1/2^-$ with varying quark masses is non-trivial. In
addition, one can also calculate the finite volume energy levels. One finds that
there are several strongly coupled $S$-wave thresholds
that induce a similar avoided level crossing as narrow resonances.
Consequently, one has to be extremely careful in comparing lattice results at
unphysical quark masses when a strong coupled channel dynamics is present.
For more details, we refer the reader to Ref.~\cite{Doring:2013glu}.
Finally, we note that pion-nucleon scattering in the negative parity channel 
in lattice QCD was considered by Lang and Verduci~\cite{Lang:2012db}.

\end{document}